\documentclass[12pt,preprint]{aastex}
\slugcomment{Submitted to The Astrophysical Journal}  
\shorttitle{Bar Diagnostics in Edge-On Spiral Galaxies. III.}
\shortauthors{Bureau \& Athanassoula}
\newcommand{\twoone}{$2\!:\!1$}
\newcommand{\fourone}{$4\!:\!1$}
\received{2004 March 8}
\begin{document}
%
%
\title{Bar Diagnostics in Edge-On Spiral Galaxies. III. \\N-Body
 Simulations of Disks}
\author{M.\ Bureau\altaffilmark{1}\altaffilmark{2}}
\affil{Columbia Astrophysics Laboratory, 550~West 120th~Street, 1027
 Pupin Hall, \\MC~5247, New York, NY~10027}
\email{bureau@astro.columbia.edu}
\altaffiltext{1}{Hubble Fellow}
\altaffiltext{2}{Now at Sub-Department of Astrophysics, University of Oxford,
 Denys Wilkinson Building, Keble Road, Oxford OX1~3RH, United Kingdom}
\and 
\author{E.\ Athanassoula}
\affil{Observatoire de Marseille, 2 place Le Verrier, F-13248 Marseille
 Cedex~4, France}
%
%
\begin{abstract} 
Present in over $45\%$ of local spirals, boxy and peanut-shaped bulges
are generally interpreted as edge-on bars and may represent a key
phase in bar evolution. Aiming to test such claims, the kinematic
properties of self-consistent 3D N-body simulations of bar-unstable
disks are studied. Using Gauss-Hermite polynomials to describe the
major-axis stellar kinematics, a number of characteristic bar
signatures are identified in edge-on disks: 1) a major-axis light
profile with a quasi-exponential central peak and a plateau at
moderate radii (Freeman Type~II profile); 2) a ``double-hump''
rotation curve; 3) a sometime flat central velocity dispersion peak
with a plateau at moderate radii and occasional local central minimum
and secondary peak; 4) an $h_3-V$ correlation over the projected bar
length. All those kinematic features are spatially correlated and can
easily be understood from the orbital structure of barred disks. They
thus provide a reliable and easy-to-use tool to identify edge-on
bars. Interestingly, they are all produced without dissipation and are
increasingly realized to be common in spirals, lending support to
bar-driven evolution scenarios for bulge formation. So called
``figure-of-eight'' position-velocity diagrams are never observed, as
expected for realistic orbital configurations. Although not uniquely
related to triaxiality, line-of-sight velocity distributions with a
high velocity tail (i.e.\ an $h_3-V$ correlation) appear as
particularly promising tracers of bars. The stellar kinematic features
identified grow in strength as the bar evolves and vary little for
small inclination variations. Many can be used to trace the bar
length. Comparisons with observations are encouraging and support the
view that boxy and peanut-shaped bulges are simply thick bars viewed
edge-on.
\end{abstract}
%
%
\section{Introduction\label{sec:intro}}
Three-dimensional (3D) N-body simulations of bar-unstable disks have
consistently shown that, soon after a bar forms, it buckles and
settles with an increased thickness and velocity dispersion, appearing
boxy or peanut-shaped (B/PS) when viewed edge-on
\citep[e.g.][]{cs81,cdfp90,rsjk91}. This is particularly important in
view of the large number of vertically extended structures commonly
refered to as B/PS bulges in edge-on spirals. The recent survey of
\citet*{ldp00a} reveals that at least 45\% of bulges across all
morphological types are B/PS \citep[but see also the earlier studies
of][]{j86,sa87,s87}, suggesting that a significant fraction of bulges
may form through dynamical instabilities in disks rather than
dissipational collapse \citep*[e.g.][]{els62} or accretion of smaller
systems \citep[e.g.][]{sz78}.

Verifying such claims is hard, as reliably identifying bars in edge-on
systems is difficult. A plateau in the major-axis light profile of an
edge-on spiral has long been claimed to indicate the presence of a bar
\citep*[e.g.][]{cc87,hw89,ldp00b}, but axisymmetric features could
equally give rise to it and end-on bars would likely remain
undetected. A kinematic identification is clearly called for,
analogous to the use of longitude-velocity diagrams in the Galaxy
(\citealt{p75,ml86,bgsbu91}; etc). This was first proposed and used in
the context of external bulges by \citet{km95}, while
\citeauthor*{ba99} (\citeyear{ba99}, hereafter \citeauthor{ba99}) and
\citeauthor*{ab99} (\citeyear{ab99}, hereafter \citeauthor{ab99})
refined those kinematic bar diagnostics using, respectively, periodic
orbit calculations and hydrodynamical simulations. These were then
applied to relatively large samples of galaxies by \citet{mk99} and
\citet{bf99}, who successfully showed a close relationship between
B/PS structures and bars. Using the ionized-gas kinematics, these
works were however only able to probe the galactic potentials in the
equatorial plane and were restricted to intermediate and late-type
spirals.

Using N-body simulations, we develop in this paper fully
self-consistent stellar kinematic bar diagnostics for edge-on
systems. Those diagnostics are most easily applicable to gas-poor
early-type spirals (where the bulges are large and contain little
dust, perfect for photometric studies), and they allow to probe
galactic potentials out of the disk plane (to test the barred nature
of B/PS structures at large galactic heights). The diagnostics rely
exclusively on readily observable quantities and are thus not meant as
a study of barred N-body models. They have already been successfully
applied by \citet{cb04} to the complete sample of \citet{bf99}, and
they are being applied at different heights to a subset of the galaxies
by Zamojski et al.\ (2005, in preparation). \citet{baabdf05} and
\citet*{aab05} present $K$-band observations of the same sample,
quantifying the B/PS structure. The goal of all those studies is to
study the vertical structure of bars, and ultimately to clarify their
relationship to bulges in the context of bar-driven evolution
scenarios \citep[e.g.][]{pn90,fb93,fb95,a03a}.

We review existing bar diagnostics for edge-on disks in
\S~\ref{sec:past_diagn} and describe the N-body simulations used
throughout this paper in \S~\ref{sec:nbody}. The basic stellar
kinematic bar diagnostics are presented in \S~\ref{sec:bar_diagn}
along with their viewing angle dependence. \S~\ref{sec:time_evol} and
\S~\ref{sec:incl_dep} describe, respectively, the time evolution of
the diagnostics (as the bar evolves) and their inclination dependence
(for galaxies close to but not perfectly edge-on). We discuss the
uniqueness of the bar diagnostics and compare them to existing
observations in \S~\ref{sec:discussion}, and our results are
summarized in \S~\ref{sec:conclusions}.
%
%
\section{Past Diagnostics\label{sec:past_diagn}}
\citet{km95} first proposed to use the major-axis position-velocity
diagrams (PVDs) of edge-on disks to detect bars in external galaxies
\citep[see also][]{k96}. PVDs show the density or luminosity of
material as a function of projected radius and line-of-sight velocity,
and they are easily obtained by long-slit (or integral-field)
spectroscopy. Although the bar signatures they predicted were similar
to those observed, the diagnostics turned out to be flawed (see
\citeauthor{ab99}; \S~\ref{sec:orb_pop}), and we focus here on the
improved diagnostics presented in \citeauthor{ba99} and
\citeauthor{ab99}.

In \citeauthor{ba99}, we used the periodic orbits in a standard family
of barred disk potentials as building blocks to model real galaxies,
using a simple recipe to populate them. The global structure of the
corresponding PVDs, obtained for reasonable combinations of orbit
families and different viewing angles, provides in itself a reliable
bar diagnostic for edge-on disks. Particularly, gaps between the
signatures of the different orbit families (as well as material in the
so-called forbidden quadrants) follow directly from the inhomogeneous
distribution of the orbits. The shape of the signature of the $x_1$
and $x_2$ orbit families constrains the viewing angle to the bar and,
to a lesser extent, the mass distribution of the bar and disk.

In \citeauthor{ab99}, the hydrodynamical simulations of \citet{a92b}
were used to study the gaseous PVDs of edge-on barred disks for
various viewing angles and mass distributions, using the same family
of potentials as for the orbits. Because of shocks and inflow in the
bar region, a characteristic gap develops in the PVDs between the
signature of the nuclear spiral (when an inner Lindblad resonance is
present) and that of the outer disk. This is a clear bar diagnostic
and the viewing angle is further constrained by the signature of the
nuclear spiral. Those kinematic bar diagnostics were successfully
applied by \citet{mk99} and \citet{bf99}, but a continuous worry is
whether the tracer used (e.g.\ H$\alpha$, [\ion{N}{2}], CO,
\ion{H}{1}, etc) effectively samples the equatorial plane.

Although they are indicative of the gaseous and stellar structures to
be expected from the PVDs of real galaxies, the above models are
clearly not fully satisfactory. They are not self-consistent, and
\citeauthor{ba99} relies solely on periodic orbits, while regular
(trapped) and chaotic orbits are expected to be
important. Furthermore, both models are two-dimensional (2D) and are
thus confined to the equatorial plane of the galaxies. This is not a
major obstacle for the gas, which typically has a small scaleheight,
but it is an unnecessary restriction for the stars, in particular if
one wants to probe the potential to large galactic heights (as for
B/PS structures) or non-exactly edge-on systems.
%
%
\section{N-Body Models\label{sec:nbody}}
\subsection{Simulations\label{sec:init_cond}}
Naturally, the next step is to develop more reliable and realistic
kinematic diagnostics for the stars, using self-consistent 3D N-body
simulations. For this, we use N-body simulations similar to those
discussed in \citet{am02}. A large number of simulations were run and
analyzed but only three representative cases are discussed here in
details: weakly, intermediate, and strongly barred disks. They are
shown in Figure~\ref{fig:strength} along with a simulation preserving
axisymmetry.

Because a number of parameters influence the strength of a bar (e.g.\
the bar mass, length, axial ratio, radial density profile, etc), an
exhaustive discussion of the whole of parameter space would be
extremely lengthy and is in any case not necessary for our goals. The
sequence chosen is thus meant to represent a growing influence of the
bar on the entire disk, as reflected for example by its vertical
extent, central concentration, and the contrast of the inner ring and
outer spiral arms (or equivalently the depletion of material around
the Lagrange points $L_{\rm 4}$ and $L_{\rm 5}$). The increasing bar
strength in the simulations discussed is illustrated in
Figure~\ref{fig:fourier}, which shows the importance of the even terms
(compared to the axisymmetric term) in a Fourier decomposition of the
face-on density distribution of each model \citep[see also][]{am02}.

The initial conditions were set roughly following \citet{h93} and
\citet{am02}, and the simulations contain only a luminous disk and
dark halo. No bulge is present, so all the quantities discussed in
this paper refer exclusively to disk material, although as we will see
much of it does acquire a large vertical extent.

The initial disk density is given by
\begin{equation}
\rho_{\rm d}(R, z)=\frac{M_{\rm d}}{4\pi h^2z_0}\,\,\exp(-R/h)\,\,{\rm sech}^2(\frac{z}{z_0}),
\end{equation}
where $M_{\rm d}$ is the total disk mass, $h$ the disk radial
scalelength, $z_0$ the disk vertical scaleheight, and $R$ and $z$ are
respectively the cylindrical radius and height. For all barred
simulations, $M_{\rm d}=1$, $h=1$, and $z_0=0.2$. The $Q$ value is
initially set to be constant throughout the disk at a value near
unity. For the three barred simulations shown in
Figure~\ref{fig:strength}, $Q=1.4$, $1$, and $1$ (from left to right).

The initial dark halo density is given by
\begin{equation}
\rho_{\rm h}(r)=\frac{M_{\rm h}}{2\pi^{3/2}}\,\frac{\alpha}{r_{\rm
    c}}\,\,\frac{\exp(-r^2/r_{\rm c}^2)}{r^2+\gamma^2},
\end{equation}
where $M_{\rm h}$ is the total halo mass, $r_{\rm c}$ and $\gamma$ are
halo scalelengths, $r$ is the spherical radius, and $\alpha$ is a
normalization constant. For all barred simulations, $M_{\rm h}=5$ and
$r_{\rm c}=10$. In order to obtain bars of different strengths, we
have followed the precepts of \citet{am02} and \citet{a02,a03a} and
used halos with adequate parameters. For the strongest bar,
$\gamma=0.5$. This halo is relatively concentrated and has substantial
mass in the region where the main resonances develop. This leads to a
large angular momentum exchange between the halo and disk component
and to a strong bar \citep{a02,a03a}. For the intermediate bar case,
$\gamma=3$, decreasing the halo mass in the resonant regions but
ensuring sufficient angular momentum exchange to form a bar of
intermediate strength. For the weak bar we ensured that the halo
resonances could not absorb much angular momentum by rendering the
material there quite hot. In practice, this was achieved as in
\citet{a03a}, i.e.\ by using an extended halo with twice the mass (see
\citealt{a03a} for details).

The left column in Figure~\ref{fig:strength} is included for
comparison and shows a simulation in which the disk does not develop a
bar. We used an artifact to achieve such stability, namely a rigid
halo that by necessity can not exchange angular momentum with the
disk. In this case, $M_{\rm d}=0.3$, $h=1$, $z_0=0.2$, and
$Q=0.1$. The halo parameters are $M_{\rm h}=5$, $r_{\rm c}=10$, and
$\gamma=0.5$. The disk thus remains stable, despite the fact that it
is cold.

In all simulations, $2\times10^5$ particles are used for the disk,
resulting in typically $0.9-1.0\times10^6$ particles for the halo. As
shown by \citet{am02} and \citet{a03a}, angular momentum exchange
between the disk and halo ensures that all barred simulations are more
centrally concentrated and are luminous matter (i.e.\ disk) dominated
in the inner parts at late times, no matter what the initial halo
concentration is. The important characteristic of the three barred
simulations selected is thus truly the strength of the bar at late
times. Our analysis was actually carried out for a much larger sample
of simulations than shown here (well over $100$), but all of them
possess features similar to those described in
\S~\ref{sec:bar_diagn}--\ref{sec:incl_dep}, although of course the
features' amplitudes vary from case to case. The following parameter
space was carefully (but not systematically) explored: $Q=0.1-2$,
$\gamma=0.5-5$, and $z_0=0.1-0.2$ (all other parameters were generally
kept fixed).

All simulations were run on a Marseille Observatory GRAPE-5 system
with a GRAPE treecode similar to that described in \citet{ablm98}. An
opening angle of $0.6$ was used with a softening length of $0.0625$
and a time step of $0.015625$, resulting in an energy conservation
better than or equal to $0.1\%$ over the entire duration of the
simulations (which are stopped at $t=900$). Full information on all
particles is saved every $20$ time units (see
\S~\ref{sec:time_evol}). Throughout the paper, the units used are such
that $M_{\rm d}=1$, $h=1$, and $G=1$. Thus, for a disk of mass
$5\times10^{10}$~M$_{\sun}$ and scalelength $3.5$~kpc, the unit of
mass is $5\times10^{10}$~M$_{\sun}$, the unit of length is $3.5$~kpc,
the unit of velocity is $248$~km~s$^{-1}$, and the unit of time is
$1.4\times10^7$~yr. To create FITS (Flexible Image Transport System)
files (see below), we use a binning of $0.08$ per pixel spatially and
$0.125$ in velocity. For the scalings above, this corresponds to
$280$~pc ($2\farcs9$) per pixel at a distance of $20$~Mpc and a
two-pixel spectral resolution of $62$~km~s$^{-1}$ ($R=4835$), similar
to the typical set-ups used in stellar kinematic studies of nearby
galaxies. To ensure a sufficient signal-to-noise ratio $S/N$ in the
outer parts when extracting the kinematics, we must use a slit width
of $0.5$, much thicker than normally used. We have verified that this
make no qualitative and minimal quantitative differences to our
results. We are however limited in our ability to derive the
kinematics at large galactic heights, where the density drops rapidly,
and a few simulations with greater numbers of particles will be
discussed at large $z$ in Zamojski et al.\ (2005, in preparation),
along with relevant observations. We thus limit ourselves here to the
major-axis kinematics.

The simulations are manipulated with NEMO \citep[e.g.][]{t95} and
folded about the equatorial plane and under an $180\degr$ rotation to
increases the $S/N$ (analogous to assuming vertical and
bi-symmetry). Long-slit spectra along the major-axis are then
extracted into FITS files for further kinematic analysis, taking into
account the necessary line-of-sight integrations. Those spectra are
fed to the XSAURON data analysis software \citep{betal01,zetal02},
where the line-of-sight velocity distribution (LOSVD) at each position
is fit with a Gauss-Hermite series. This is necessary because, as will
be shown in \S~\ref{sec:bar_diagn}, the LOSVDs often deviate
significantly from a Gaussian and much information is contained in the
high and low velocity wings \citep[see][]{mf93,g93}. It also allows us
to stick as closely as possible to observations, where absorption line
(i.e.\ stellar kinematic) data are normally parametrized by
Gauss-Hermite series. Although this requires high $S/N$, reliable
profiles are now routinely available in the literature (at least up to
$h_3$; e.g.\ \citealt{bsg94,f97,cb04,eetal04}). The mean velocity $V$
and velocity dispersion $\sigma$ can be chosen to have their usual
meanings, while the third ($h_3$) and fourth ($h_4$) order terms
represent respectively the asymmetric (skewness) and symmetric
(kurtosis) departures from a pure Gaussian. Clearly, however, the
deconvolution step of observational data analysis (arguably the most
sensitive) is neither necessary nor possible with N-body
data. Furthermore, any comparison with real data must intrinsically
assume a certain mass-to-light ratio $M/L$ for all luminous N-body
particles, which we take here as constant. This is probably a
satisfactory assumption in the inner parts of galaxies
\citep[e.g.][]{k86,pb96} and at near-infrared wavelengths.
\subsection{Relevance and Limitations\label{sec:limitations}}
Both the initial conditions and the assumptions built in a pure N-body
treatment are of course highly idealized. This holds not only for the
simulations discussed here, but also for the vast majority of N-body
simulations presented so far in the literature. In particular, such
simulations start with quiet axisymmetric disks, with specific mass,
velocity, and velocity dispersion radial profiles, and they do not
include a gasous component. We will briefly discuss those assumptions
here, before turning to a comparison of simulated and real bars. 

Our initial conditions assume a ${\rm sech}^2$ law for the vertical
disk density and a halo with a core. We ran other simulations with an
initial vertical slab geometry, with similar results. \citet{am02} and
\citet{a02,a03a} also clearly showed that, even for highly
concentrated and/or dominant halos, luminous matter can dominate the
inner parts of the models at late times. In fact, contrary to accepted
wisdom, and as long as they are relatively cold, massive halos
ultimately lead to stronger bars. Only few experimentations with a
cosmologically-motivated (but still idealized) halo \citep*{nfw96}
have been reported so far \citep[e.g.][]{vk03,mw04}. The results of
\citet{vk03} are very similar to those of \citet{a03a} with an
extended halo, particularly regarding the bar length. Nevertheless,
more thorough and quantitative comparisons are clearly required to
fully assess the effect of cosmologically motivated halos on bar
structure. The effect an initial bulge would have on disk particles is
very much like that of a dark halo, although the bulge also directly
affects observables. In particular, because of the superposition of
the bulge, the apparent strength of any boxy or peanut shape resulting
from disk thickening decreases \citep{a05}. Some simulations with an
initially radially varying $Q$ parameter are discussed by
\citet{a03a}, but no difference liable to change any of the results of
the current paper was found.

\citet{bhsf98} have shown that the presence of gas in simulations
tends to weaken both the bar and the B/PS bulge resulting from disk
thickening. A substantial amount of gas is however required, making
the process irrelevant for early and intermediate type spirals, and
the central masses resulting form gaseous inflow are rather large. In
any case, clear B/PS bulges are observed in gas-rich late-type
galaxies \citep{ldp00a}, questioning the true efficiency of such
processes. Total bar disruption \citep[e.g.][]{pn90,fb93} is also in
conflict with the large fraction of bars observed, unless they can be
regenerated rapidly, which requires substantial external gas accretion
\citep[e.g.][]{bc02}. The latter issue is still open observationally.

As for non-idealized initial conditions extracted directly from
cosmological simulations of structure formation, while those
simulations often include gas and simple recipes for star formation
and evolution, the spatial and mass resolutions are usually too crude
to study the dynamics of the systems in details. Thus, very few
results on the dynamical evolution of disk substructures have been
obtained by such simulations, in contrast to the wealth of results
obtained with idealized models.

To summarize, while our simulations do have shortcomings, they allow a
robust analysis and a well-grounded discussion. In particular, since
many of the kinematic bar properties identified in this paper can be
related to generic properties of orbits (see below), we believe that
most of our results should also be generic.

As for a detailed comparison of our simulations (and similar ones)
with real galaxies, one might think that our strong bar case seen
nearly side-on yields a B/PS bulge more extreme than commonly observed
in nearby galaxies (see, e.g., Fig.~\ref{fig:strength}). However, a
detailed look at a large sample of edge-on galaxies shows that this is
not the case. While there are no statistics on the incidence of
"extreme" peanut-shaped bulges, \citet{ldp00a} found that about $4\%$
of all (edge-on) galaxies have a clear peanut-shape bulge ($9\%$ of
all non-spheroidal bulges) and $20\%$ have a clear box {\em or}
peanut-shape bulge ($44\%$ of all non-spheroidal bulges). Those
fractions are non-negligeable considering that peanut-shaped bulges
are only observed for rather strong bars seen nearly side-on. Indeed,
as the angle between the bar minor-axis and the line-of-sight
increases, the apparent peanut strength decreases. As they are usually
shown perfectly side-on, peanut-shaped bulges in simulations tend to
appear more extreme than observed peanuts, which have various
orientations. Perhaps most important, as mentioned above, the presence
of a classical bulge in addition to a bar also decreases the apparent
strength of the peanut toward a more boxy shape \cite[e.g.][]{a05}.

Some well-known extreme examples of B/PS bulges include
\objectname{IC~4767}, first studied in detail by \citet{wb88}, but a
very strong peanut-shaped bulge is also clearly seen in a number of
objects from the \citet{bf99} and \citet{cb04} sample, even in their
poor quality optical Digitized Sky Survey (DSS) images. Examples
include \objectname{ESO151-~G004}, \objectname{NGC~2788A},
\objectname{ESO443-~G042}, \objectname{NGC~6771},
\objectname{ESO597-~G036}, and a number of weaker cases. The bulges'
shapes are in fact even more extreme at $K$-band, generally considered
a better tracer of the mass in spiral galaxies and unaffected by dust,
as will be shown in \citet{baabdf05} and \citet{aab05}. Better spatial
resolution also usually yields sharper B/PS features, as is clearly
illustrated by the Hubble Heritage image of
\object{ESO597-~G036}\footnote{http://heritage.stsci.edu/1999/31/}.

Other observational properties of B/PS bulges also tend to compare
favorably with our or similar N-body simulations. No thorough
comparative study exists, but a number of observables have been
measured in both observations and simulations and several comparisons
of specific properties have been made. \citet{am02} and \citet{a03b}
make a strong case that the behavior of their barred N-body models
agrees well with observations of face-on barred galaxies, for
quantities as varied as Fourier amplitudes of the mass distributions,
surface brightness profiles, axial ratios, and isophotal
shapes. \citet{a05} extend those comparisons to edge-on systems, in
particular using surface brightness cuts, median-filtered images, and
velocity fields. A positive comparison of the kinematic properties of
edge-on galaxies and the current simulations is provided by
\citet{cb04} and in \S~\ref{sec:observations}.
%
%
\section{Bar Diagnostics\label{sec:bar_diagn}}
In addition to the model with no bar, Figure~\ref{fig:strength} shows
the stellar morphology and major-axis kinematics of the three selected
simulations with a weakly, intermediate, and strongly barred disk, all
viewed edge-on. The bars appear rather round when seen end-on
(line-of-sight parallel to the bar), boxy or slightly peanut-shaped
when viewed at intermediate angles, and strongly peanut-shaped when
seen side-on (line-of-sight perpendicular to the bar; see
\citealt{cs81,cdfp90,rsjk91}). All the bars could easily be confused
with a ``classical'' bulge (i.e.\ a nearly axisymmetric, vertically
extended structure containing no disk material) based on the
photometry alone, although the simulations contain no such
material. To provide the tools to discriminate between the two, we
focus here on identifying characteristic signatures of triaxiality in
the (projected) stellar kinematics.

\placefigure{fig:strength}

\subsection{Position-Velocity Diagrams\label{sec:PVD}}

Although observed stellar PVDs are usually rather noisy, prompting a
parametric description such as the Gauss-Hermite fits used here, the
PVDs of Figure~\ref{fig:strength} contain a wealth of information
which can be interpreted directly in orbital terms. For this, we rely
heavily on the discussion of PVDs from individual periodic orbit
families presented in \citeauthor{ba99}, even though they were in 2D
only (see \citealt{c80} and \citealt{cg89} for a more extensive
descriptions of 2D orbit families). The orbit families in the N-body
case can be viewed as a 3D generalization of those, and for simplicity
we will hereby refer to the ``trees'' of orbit families related to the
$x_1$ (elongated parallel to the bar) and $x_2$ (elongated
perpendicular to the bar) orbits simply as $x_1$ and $x_2$ \citep[see,
e.g.,][]{spa02a,spa02b}.

The PVDs of the axisymmetric case do not vary with viewing angle and
their upper envelopes roughly follow the circular velocity curve, with
a characteristic tail of low-velocity material ($h_3-V$
anti-correlation) due to the projected outer disk. The PVDs of the
strong bar case most clearly delineate the signatures of the different
barred orbit families. Two families dominate. First, the $x_1$ orbits
give rise to the central parallelogram-shaped feature in the PVDs (see
Fig.~\ref{fig:strength}). This particular shape arises because their
axial ratio $a/b$ increases with decreasing radius ($a$ and $b$ are,
respectively, the semi-major and semi-minor axes), and the inner $x_1$
orbits reach correspondingly higher (along the major-axis) or lower
(along the minor-axis) velocities compared to the outer ones
(\citeauthor{ba99}). Thus, when the bar is seen end-on, the
parallelogram-shaped signature of the $x_1$ orbits reaches high
line-of-sight velocities (particularly in the center) and has a small
projected width. It thus appears thin (long and narrow). When the bar
is seen side-on, the parallelogram-shaped signature of the $x_1$
orbits reaches only low projected velocities and has a large projected
width, appearing fat (short and wide). Generally however, the N-body
signature of the $x_1$ orbits is wider than that predicted in
\citeauthor{ba99}, particularly when the bar is seen exactly end-on or
side-on. This is expected since the orbits in the simulations are
regular, i.e.\ trapped around stable periodic orbits, or chaotic. The
current PVDs can thus be thought of as blurred versions of those in
\citeauthor{ba99}, in the same manner as the blurred periodic orbit
distributions of \citet{psa02} should roughly resemble real galaxies.

The second dominant feature of the PVDs is the bright almost
solid-body feature which crosses the $x_1$ parallelogram in the
central parts and becomes flat in the outer parts
(Fig.~\ref{fig:strength}). It is partly due to the orbits composing
the inner ring (high order families within corotation; see
\citealt{psa03b}), but mostly to the multitude of stars and orbits
beyond corotation (and seen in projection). The parent orbits are thus
close to circles and their signature in the PVDs does not vary much
with viewing angle. Because the orbits are not perfectly circular,
however, their signature is not exactly solid-body but slightly curved
(e.g.\ outer \twoone\ orbits; see, again, \citeauthor{ba99}).

It is interesting to note that we do not see much evidence for $x_2$
orbits. These should be most obvious in the PVDs when the bar is seen
side-on (\citeauthor{ba99}), but they are not observed (at least not
in large numbers).

The superposition of the outer disk signature to that of the $x_1$
orbits (which is viewing angle dependent) is at the origin of most of
the bar features observed in the PVDs, and thus in the Gauss-Hermite
kinematic profiles (Fig.~\ref{fig:strength}). As such, the latter can
reliably be used both to identify the presence of a bar and to
constrain its viewing angle. In particular, the $x_1$ orbits appear
responsible for the tail of high velocity material ($h_3-V$
correlation) observed at most viewing angles, which we argue is a good
indicator of triaxiality. Those effects are discussed in details
below.

\subsection{Surface Brightness Profiles\label{sec:I}}

Although not strictly speaking a kinematic quantity, the surface
brightness profile along the major-axis can also be derived from
kinematic observations, for example by summing long-slit data in
wavelength. For the axisymmetric case, the surface brightness roughly
follows an exponential profile, as expected from the initial
conditions. For the strong bar case seen end-on, the bar appears as a
prominent central peak in the light profile, with an approximately
exponential but steep decline (see Fig.~\ref{fig:strength}). A local
minimum is created at moderate radii, followed by a slight rise up to
the end of the bar (the inner ring; $X\approx4.0$) and a smooth
decline at larger radii. The inner ring is easily visible in both the
strong and intermediate bar cases. The local minimum becomes a broad
``shoulder'' at intermediate viewing angles and a truly flat plateau
when the bar is seen side-on. This plateau can be identified with the
plateaus often claimed to trace bars in the edge-on galaxy photometry
literature \citep[e.g.][]{cc87,hw89,ldp00b}. The width of the central
peak also increases slightly with viewing angle. Both the central peak
and the shoulders/plateaus are still relatively strong in the
intermediate bar case, but only the central peak remains prominent for
the weak bar. This peak is testimony that a substantial radial
rearrangement of material occurs, even for the weakest bars (see
\S~\ref{sec:time_evol}).

A prominent and steep central peak with quasi-exponential light
profile thus appears characteristic of bars in our models, although it
would normally be associated with a classical bulge in standard
bulge-disk decompositions \citep*[e.g.][]{apb95,j96,mch03}.

\subsection{Velocity Profiles\label{sec:V}}

As expected from the initial circular velocity curve of the model, the
rotation curve of the axisymmetric case has a rapid but smooth rise
and remains flat at large radii. As shown by the strongly barred case
of Figure~\ref{fig:strength}, however, the rotation curve of a barred
disk seen end-on has a strong ``double-hump'' structure. That is, the
rotation curve first rises rapidly and reaches a local maximum, it
then drops slightly and creates a local minimum, and it rises again
slowly up to its flat section. This behavior is weakened at
intermediate viewing angles, when the local maximum and minimum
disappear to form a single plateau at moderate radii. Once seen
side-on, the disk rotation curve appears almost completely solid-body,
with a steeper gradient just before the flat part. As expected, the
double-hump structure is weakened for the intermediate bar case, where
only broad plateaus are visible at moderate radii for all viewing
angles. For the weak bar case, the double-hump structure has
essentially completely disappeared.

For the barred cases, we note that the velocity reached by the rapidly
rising part of the rotation curve (the first hump) decreases with
increasing viewing angle, suggesting that this feature is caused by
the (mostly inner) $x_1$ orbits. The radius at which the rotation
curve becomes flat simultaneously increases, and it appears to mark
the beginning of the inner ring, which is slightly elongated parallel
to the bar \citep[e.g.][]{am02}. In fact, the flat part of the
rotation curve is usually reached at the end of the plateau in the
light profile, while the local velocity minimum at moderate radii (or
more generally the end of the first velocity plateau) occurs roughly
at the end of the central luminosity peak (see \S~\ref{sec:I}).

Whether or not there is a local maximum and minimum at moderate radii,
a double-hump structure in the rotation curve appears characteristic
of bars viewed edge-on (except for the weakest cases), although as we
will argue below (\S~\ref{sec:h3-v}) it is probably not uniquely
related to them.

\subsection{Velocity Dispersion Profiles\label{sec:sigma}}

The velocity dispersion profile of the axisymmetric case is unusual
but simply reflects the fact that the disk is very cold, the
dispersion mainly resulting from the line-of-sight integration through
the outer disk (the observed $\sigma$ should in principle go to zero
in the center for a perfectly cold disk). The velocity dispersion
profile of the strong bar case seen end-on is also
characteristic. Rather than a sharp central peak, the profile has a
broad and rather flat central maximum followed by a sharp drop and a
significant secondary (local) maximum. At intermediate and large
viewing angles, the central peak is smaller, its decrement shallower,
and the secondary maximum more important, such that a broad shoulder
or plateau is formed at moderate radii (although a clear secondary
maximum is still visible when the bar is seen side-on). For the
intermediate bar case, only a narrow central peak is present, without
a flat section, and the secondary peak is rather weak, again forming a
shoulder or plateau at moderate radii. The viewing angle dependence is
also much weaker. For the weakest bar case, only a central peak
superposed on a much broader and shallow increment is present for all
viewing angles. We note that those dispersion features were also seen
by \citet{am02}, where they are much stronger because only a thin
strip of particles around the major-axis was considered (no
line-of-sight integration).

At least for the strong bar case, the width of the central flat
section of the dispersion profile is always roughly the same as that
of the rapidly rising part of the rotation curve, suggesting that they
have a common origin ($x_1$ orbits; see \S~\ref{sec:V}). The extent of
the central $\sigma$ peak is also aproximately equal to that of the
central light peak (\S~\ref{sec:I}). Finally, the secondary dispersion
maximum always occurs just within the flat part of the rotation curve,
suggesting that it has its origin towards the end of the bar, just
inside the inner ring. This is most likely due to the tips of the last
$x_1$ orbits and the inner \fourone\ orbits. In particular, $x_1$
orbits with loops around the major-axis are expected in strong bars
\citep[e.g.][]{a92a}, causing a local increase in the velocity
dispersion, strongest for bars seen side-on. The higher energy inner
\fourone\ orbits have similar loops, which will increase $\sigma$ for
bars seen both side-on and end-on. Higher order orbit families within
corotation may increase the dispersion somewhat (at all viewing
angles), but they will mainly contribute to the excess surface density
in the inner ring \citep{psa03a,psa03b}.

We also note that for a strong bar, a large variety of orbital shapes
is encountered, particularly along the major-axis. This leads to an
increase in the observed velocity dispersion compared to face-on
views, particularly when the bar is seen end-on. For weak bars or
ovals, where the orbits are almost self-similar (i.e.\ concentric),
the orbital variety is greatly diminished, as is the viewing angle
dependance. The small local central velocity dispersion minimum
observed for strong end-on bars will be discussed further in
\S~\ref{sec:sig_min}.

Generally speaking, thus, one should expect a (sometime broad and
rather flat) $\sigma$ peak with a broad shoulder and/or a secondary
maximum as the characteristic signature of a (strong) bar viewed
edge-on. The velocity dispersion profile is also highly correlated
with the major-axis surface brightness profile.

\subsection{$h_3$ Profiles\label{sec:h3}}

As the PVDs of Figure~\ref{fig:strength} clearly show, the LOSVDs have
a complex shape in the bar region, and the Gauss-Hermite terms $h_3$
and $h_4$ are necessary to provide a good description. For the
axisymmetric case, $h_3$ is essentially always anti-correlated with
$V$, as expected. For the strong bar case seen end-on, however, $h_3$
is correlated with $V$ over the entire projected bar length, that is
until the secondary maximum in $\sigma$ or just inside the flat part
of the rotation curve. At larger radii, the $h_3$ profile is
anti-correlated with $V$ for some distance before being correlated
again. As the viewing angle is increased, the nature of the
correlations does not change, but they become much weaker. For a bar
seen side-on, the $h_3$ profile is almost flat. The $h_3-V$
correlation over the projected bar length is relatively strong for the
intermediate bar case and is greatly diminished but still present for
the weak bar, which makes it a very good tracer of bars viewed
edge-on.

The correlation of $h_3$ and $V$ over the bar region is most likely a
key tracer of triaxiality, although it is not uniquely related to it
(see \S~\ref{sec:h3-v}). Indeed, for an axisymmetric disk viewed
edge-on (or a slowly rotating spheroid), one would generally expect
$h_3$ and $V$ to be anti-correlated, since projection effects
systematically create a tail of low-velocity material at each
position. Only elongated motions (such as those of the $x_1$,
\fourone, etc orbits) can create a tail of high velocity material, as
required for $h_3$ and $V$ to correlate. The viewing angle dependence
of the $h_3-V$ correlation can thus be traced to the viewing angle
dependence of the (parallelogram-shaped) $x_1$ orbits signature in the
PVDs (see \S~\ref{sec:PVD}; \citeauthor{ba99}).

For barred galaxies viewed exactly edge-on (such as the simulations of
Fig.~\ref{fig:strength}), the $h_3-V$ correlation is strengthened by
the superposition of the outer disk signature to that of the $x_1$
orbits, which lowers the mean velocity (at all positions) and
correspondingly ``magnifies'' the high velocity tail. The $h_3-V$
correlation is then expected to be stronger if there is less material
in the forbidden quadrants, which probably explains why the weak bar
case still shows a significant signature (see, e.g., the $a/b$
sequence in \citeauthor{ab99}).

The situation is more complex for slightly inclined systems, where the
outer disk is not necessarily seen (in projection), and for hot
systems such as spheroids, where significant pressure support may be
present. We discuss those issues in more details in
\S~\ref{sec:incl_dep} and \ref{sec:h3-v}, respectively. Nevertheless,
generally speaking, a correlation of $h_3$ and $V$ appears
characteristic of bars viewed edge-on.

\subsection{$h_4$ Profiles\label{sec:h4}}

Although $h_4$ is hard to measure observationally, we discuss it here
for completeness. For the strong bar seen end-on, a strong and rather
flat minimum is present in the center, followed by a sharp rise and a
more gentle decline at moderate radii. Those structures broaden and
weaken substantially with increasing viewing angle, such that the
$h_4$ profile is largely featureless at even moderate viewing
angles. Weak features are visible for the intermediate bar case seen
end-on, but for other viewing angles and for all viewing angles of the
weak bar case, the $h_4$ profile is essentially flat.

At least for the strong bar case, the width of the central $h_4$
minimum is roughly the same as that of the flat part of the dispersion
profile (and thus of the rising part of the rotation curve), perhaps a
bit larger. In fact, $h_4$ and $\sigma$ have roughly opposite
behaviors. It is thus likely that all three features are due to the
inner $x_1$ orbits.

\subsection{Characteristic Bar Features\label{sec:bar_features}}

From the discussion above (Fig.~\ref{fig:strength}), the
characteristic kinematic signatures of bars viewed edge-on can be
summarized as follows: 1) a quasi-exponential central peak in the
surface brightness profile, with a shoulder or plateau at moderate
radii; 2) a double-hump rotation curve, with or without a local
maximum/minimum at moderate radii; 3) a possibly broad and rather flat
central velocity dispersion peak, with a sharp edge and a shoulder or
plateau (and possibly a secondary maximum) at moderate radii; 4) a
correlation of $h_3$ and $V$ over the projected bar length (i.e.\
within the flat part of the rotation curve). A local central $\sigma$
minimum may also be present and the lengths of those features are not
arbitrary, but correlated (see \S~\ref{sec:I}--\ref{sec:h4}).

The bar signatures tend to weaken as the viewing angle increases or
the bar strength decreases, causing some degeneracy between the
two. It should thus be relatively easy to kinematically identify an
edge-on bar (unless it is very weak), but any constraint on the
viewing angle will be rough at best. This parallels the situation for
the gas (\citeauthor{ab99}) and in our Galaxy (see e.g.\ \citealt{k96}
for a review). The light profile and rotation curve seem to be the
most degenerate quantities, the velocity dispersion and $h_3$ profiles
less so. In particular, the $h_3-V$ correlation remains present even
for weak bars. Clearly, whether one can actually disentangle bar
strength from viewing angle will strongly depend on the quality of the
data.

It is also interesting to note that the kinematic bar signatures are
strongest when the bar is seen end-on and appears round, whereas it is
easiest to identify bars morphologically when seen side-on because of
the B/PS morphology. The kinematic and morphological methods are thus
very much complementary, and they should permit to identify a bar if
present in almost any edge-on system (unless it is very weak).

Lastly, it should be kept in mind the the concept of bar strength is
somewhat ill-defined. Not only can a bar be strong in the inner parts
and weak in the outer parts (or vice-versa), but many parameters
influence the dynamics (e.g.\ bar axial ratio, mass, pattern speed,
etc; e.g.\ \citealt{a92a,a92b}). The kinematic features described
above may therefore not all be present systematically in every object,
but they should be considered separately, each individual feature
yielding clues about the orbital structure.

The length of a bar is a similarly difficult quantity to measure
precisely \citep[e.g.][]{am02}, but many of the kinematic features
present in the profiles of Figure~\ref{fig:strength} can be used as
rough yardsticks. First, the end of the plateau in the major-axis
light profile and the position where the rotation curve becomes flat
both appear to trace the position of the inner ring. The secondary
$\sigma$ maximum, the end of the $h_3-V$ correlation region, and the
secondary $h_4$ minimum are also usually roughly equal although
slightly shorter than the previous two quantities (more so for $h_3$),
and they probably trace better the end of the bar itself. The exact
location of all those features also depends somewhat on the
orientation of the bar, reflecting the fact that most orbit families
as well as the inner ring are elongated
\citep[e.g.][]{am02}. Nevertheless, if the location of corotation (or
any other resonance) can be independently determined, those
measurements of the bar length will provide useful constraints on the
bar pattern speed in the usual dimensionless manner ($r_{\rm
cr}/r_{\rm bar}$, where $r_{\rm cr}$ is the corotation radius and
$r_{\rm bar}$ the radius of the bar).
%
%
\section{Time Evolution\label{sec:time_evol}}
For the strong bar case only, Figure~\ref{fig:time} shows the complete
time evolution of the various major-axis kinematic profiles ($\mu_I$,
$V$, $\sigma$, $h_3$, and $h_4$), from the initial conditions through
bar formation and buckling to the end of the simulation. The stellar
distribution changes rapidly and has high order structures ($m>2$)
early on ($t\lesssim160$). Bar formation occurs at $t\approx180-240$
and buckling at $t\approx260-320$, while significant vertical
asymmetries persist until $t\approx440$. From then on, the evolution
is smooth but continous. We look at the temporal behavior of each
kinematic quantity individually below, but after bar formation the
general trend is that, like the bar itself, most kinematic bar
signatures grow in strength and length with time. This is expected
because, as the bar lengthens, the various orbit families and their
associated signatures do so as well. The bar growth and lengthening is
most likely due to a transfer of angular momentum from the disk to the
halo and is explored further in \citet{a02,a03a}.

\placefigure{fig:time}

The moment the bar forms, the material in the disk is substantially
rearranged and a quasi-exponential central peak develops in the
surface brightness profile. For all viewing angles, the intensity of
the central peak grows with time, simultaneously with the flatness of
the plateau at moderate radii and the intensity of the secondary
maximum (if present). The extents of the central peak and plateau also
grow with time as the bar lengthens.

For the rotation curve, the central part steepens significantly during
bar formation and the double-hump structure establishes itself fairly
rapidly. Once formed, the rapidly rising part of the rotation curve
does not change much. The velocity plateau at moderate radii, however,
stretches with time and becomes an increasingly deep local
minimum. The radius at which the rotation curve becomes flat
simultaneously increases. The same is true at all viewing angles, with
the caveat that the features are generally weaker with increasing
viewing angle (see \S~\ref{sec:bar_diagn}). The characteristics bar
signatures in the velocity profile thus increase in strength and
length with time, except in the inner parts.

The central peak in the velocity dispersion profile steepens and
increases dramatically during bar formation. The local minimum and
secondary maximum at moderate radii appear soon after and their
strength and distance from the center increase with time. Again, the
evolution is roughly the same at all viewing angles although the
features are weaker for large viewing angles
(\S~\ref{sec:bar_diagn}). Interestingly, the width of the central peak
does not change much for end-on views, but it does grow slightly for
side-on views, suggesting that only the length of the bar but not its
width is increasing. The peak value increases throughout. The central
dip in the dispersion peak at $t=0$ is probably a consequence of the
approximations involved in the asymmetric drift correction for the
inner parts (when setting up the initial conditions; see
\citealt{h93}) and of the fact that the disk is initially rather cold
kinematically (see \S~\ref{sec:init_cond} and \ref{sec:sigma}). As the
local minimum disappears when the disk slushes around at early times
and the bar forms, we are confident that the $\sigma$ behavior at
later times is genuine.

The time evolution of the $h_3$ profiles is very simple. For all
viewing angles, both the central region where $h_3$ correlates with
$V$, the following $h_3-V$ anti-correlation region, and the outside
region where $h_3$ correlates with $V$ again all grow in length with
time. The gradients become slightly more shallow, however. The initial
$h_3-V$ correlation within $0.5$ disk scalelength is probably again an
artefact of the (imperfect) asymmetric drift correction at $t=0$. It
weakens at early times and disappears totally during bar formation and
buckling, only to reappear once the bar is fully formed
($t\approx320$), so the $h_3$ behavior at later times is reliable.

Analogous to the $\sigma$ evolution, the central minimum in $h_4$
grows with time, although it is truly the amplitude of the secondary
maximum which increases. Furthermore, while the width of the central
minimum remains largely unchanged as the bar evolves, the width of the
secondary maximum increases. The same behavior is observed for all
viewing angles, but the profiles are almost featureless for even
moderate ones (see \S~\ref{sec:h4}). The central maximum at $t=0$
disappears soon after the bars forms and should not affect the
subsequent bar evolution.
%
%
\section{Inclination Dependence\label{sec:incl_dep}}
It is essential to understand how the kinematic bar signatures
identified vary with the inclination $i$, for two main reasons. First,
the signatures are likely to depend sensitively on $i$, as the
line-of-sight integration through the outer disk (which acts like a
kinematic ``screen'') decreases significantly even for moderate
offsets. Second, optical observations of intermediate and late-type
spirals usually call for close to but not exactly edge-on systems, as
thick dust lanes are common and prevent the line-of-sight from
reaching the central parts of the galaxies in perfectly edge-on
objects. This is well illustrated by the work of \citet{cb04}.

Figure~\ref{fig:incl} shows the inclination dependence of the
major-axis kinematic profiles for the strong bar case. For a bar seen
end-on, the value of the central peak in the surface brightness
profile varies little with $i$, but its prominence increases as $i$
decreases (from $90\degr$, i.e.\ exactly edge-on, to lower
values). This is because the level of the plateau at moderate radii
decreases as the low density regions on the minor-axis of the bar
(around the Langrangian points $L_4$ and $L_5$) become exposed. This
effect is thus much weaker for intermediate viewing angles (although a
break appears at the end of the bar), and it is almost absent when the
bar is seen side-on.

\placefigure{fig:incl}

For an end-on bar, the central gradient in the rotation curve steepens
(and the maximum velocity reached increases) as $i$ decreases. This is
because the moderating influence of the outer disks decreases and the
$x_1$ orbits (also seen end-on) are more fully exposed. The effect is
easily seen in the PVDs of Figure~\ref{fig:incl}, where the quasi
solid-body contribution of the (projected) outer disk to the PVDs
gradually disappears as $i$ decreases, resulting in higher mean
velocities. The effect is of course opposite for large viewing angles,
where the the bar and the $x_1$ orbits are seen more side-on. There,
the central gradient in the rotation curve becomes much shallower, and
the velocity profile shows a sharp break at the end of the bar
(connecting to the flat part of the rotation curve).

The velocity dispersion profile is also strongly affected by the
gradual disappearance of the (projected) outer disk as $i$
decreases. For end-on views, although the value of the maximum does
not change, the width of the central peak widens dramatically
(essentially to the width of the bar itself) and its edges become
sharper (almost a step function). For intermediate and large viewing
angles, the width of the central peak changes little but the edges
also become sharper. In fact, the most important change is the
increase of the secondary $\sigma$ maximum at moderate radii, which
reaches values comparable to that of the central peak for ``small''
inclinations (e.g.\ $i=75\degr$). This is probably testimony to the
large vertical motions of the orbits near the end of the bar, as shown
by \citet{spa02a,spa02b} and \citet{psa02}, and to the loops and large
rectangularity of the $x_1$ and \fourone\ orbits in that region. In
fact, for side-on views, as the inclination is decreased, the slit
encompasses only material increasingly near the major-axis of the bar,
and this material is known to have a dispersion similar to that shown
in Figure~\ref{fig:incl} \citep{am02}.

Surprisingly, while for perfectly edge-on systems the $h_3-V$
correlation decreases with increasing viewing angle, the opposite is
true for objects significantly less inclined ($i\lesssim80\degr$). For
a bar seen end-on, the $h_3-V$ correlation decreases as $i$ decreases,
and there is in fact a slight anti-correlation for
$i\lesssim80\degr$. As shown by the PVDs of Figure~\ref{fig:incl},
this is because as $i$ decreases and the PVD signature of the
(projected) outer disk disappears, the mean velocities increase (see
above) and the asymmetric tail of material in the LOSVDs switches from
high to low velocities. For large viewing angles, however, when the
bar and the $x_1$ orbits are seen more side-on, the mean velocities
actually decrease with decreasing $i$ (again, see above), and the tail
of material in the LOSVDs remains at high velocities (especially
toward the ends of the bar).

For bars seen end-on, where most of the structure in the $h_4$
profile is observed (see \S~\ref{sec:bar_diagn}), the shape of the
central parts of the profile remain largely unchanged as $i$
decreases, although all the values themselves decrease. This leads to
a wide and deep central minimum with a small and sharp central peak at
small inclinations. The $h_4$ profiles are still rather featureless at
intermediate and large viewing angles, although a sharp break develops
at the end of the bar for small inclinations.
%
%
\section{Discussion\label{sec:discussion}}
\subsection{Orbit Populations\label{sec:orb_pop}}

While the PVDs of \citeauthor{ba99} were generated by selecting orbits
equally spaced along the bar minor-axis, a much better fit to the N-body
PVDs (at least for the $x_1$ orbits) is obtained by selecting orbits
with an equal increment of the Jacobi integral $E_{\rm J}$ ($E_{\rm
J}=E-\Omega_{\rm p}J_{\rm z}$ is the only conserved quantity in a
barred galaxy, where $\Omega_{\rm p}$ is the pattern speed of the bar
and $J_{\rm z}$ the angular momentum along the axis of
revolution). Such models were briefly discussed in \citeauthor{ba99}
but no PVD was shown. The characteristic curve of the $x_1$ orbits
(semi-minor axis vs.\ $E_{\rm J}$) is typically composed of two
sections: a first rather flat section where the semi-minor axis barely
changes for large increments in $E_{\rm J}$, and a second almost
vertical section where the semi-minor axis increases very rapidly for
just a small $E_{\rm J}$ increment (see, e.g., \citealt{a92a};
\citeauthor{ba99}). Our results thus imply that the first section of
the $x_1$ characteristic is preferentially populated. This is not
entirely surprising, since we have shown that a steep density gradient
develops at the center of our models (e.g.\ Fig.~\ref{fig:time}),
implying that the inner $x_1$ orbits are substantially more populated
than the outer ones.

We also note that no ``figure-of-eight'' is observed in the stellar
PVDs, under any circumstance (see
Figs.~\ref{fig:strength}--\ref{fig:incl}). This bar diagnostic was
popularized by \citet{km95} but results from a very selective
population of the periodic orbits (no overlapping or self-intersecting
orbits, particularly near corotation, and only a few outer \twoone\
orbits; see also \citealt{m96}). Our models show that no such figure
arises for realistic orbital configurations. Using simulations
undergoing analogous bar-driven evolution, a similar conclusion was
reached by \citet{od03}, but they failed to identify alternative bar
signatures. Contrary to their claim, observational techniques can be
used with $\sim10^6$ particle simulations, if the right tools are
used. As shown in \citeauthor{ab99}, a figure-of-eight feature can
develop in gaseous PVDs, but only for very strong bars, and only
extending to corotation (i.e.\ it arises from $x_2$ orbits and the
inner ring, rather than $x_1$ orbits and an outer ring as advocated by
\citealt{km95}). These results appear confirmed by observations
\citep{km95,mk99,bf99,cb04}.

\subsection{Central Light Peak and Freeman Type II
Profiles\label{sec:central_peak}}

Independently of the issue of their 3D shape, so-called secular
evolution scenarios for bulge formation have recently gained much
momentum from the realization that many (and perhaps most) bulges have
central light profiles more closely resembling a disk-like exponential
than an $R^{1/4}$ law (expected from violent relaxation), with a tight
correlation between the scalelengths of the bulge and disk (e.g.\
\citealt{apb95,j96,mch03,bgdp03}; see also \citealt{k93}). As
illustrated in Figure~\ref{fig:time}, bar formation and evolution is
accompanied by the development and continued growth of a very dense
(but still slightly elongated) central component within $1-1.5$
original disk scalelengths \citep[see also][]{am02,a02}. In our
major-axis surface brightness profiles
(Figs.~\ref{fig:strength}--\ref{fig:incl}), this component would
invariably be identified with a bulge, although there is no such
material to start with in our simulations. Certainly, the central peak
meets the fairly general definition of a bulge as the component in
excess of the inward extrapolation of the outer exponential disk
\citep*[e.g.][]{cfw99}. Morever,
Figs.~\ref{fig:strength}--\ref{fig:incl} show that the shape of the
central profiles is very close to exponential, so it seems natural to
associate these central peaks, and thus bars, with the
quasi-exponential bulges observed in more face-on systems
\citep{a05}. Although this will be more fully investigated in a future
paper \citep{aab05}, it is also interesting to note that in all but
the weakest bar cases, the central parts of the bar appear to have a
smaller scaleheight than the outer parts, leading to the
characteristic peanut shape when the bar is viewed edge-on.

Most of the (edge-on) major-axis surface brightness profiles from our
simulations would qualify as Freeman Type~II profiles \citep{f70},
having inner disks well below the inward extrapolation of the outer
exponential. Our simulations clearly show that Type~II profiles arise
from both a radial and vertical redistribution of material (and to a
lesser extent azimuthal), the stars in the barred region
simultaneously moving in radially (due to angular momentum exchange
with the halo; e.g.\ \citealt{a02,a03a}) and out vertically (due to
vertical instabilities and the creation of the B/PS bulge). Obviously,
only the first mechanism is relevant for face-on systems. The
complicated surface bightness profiles of our models therefore do not
stem from the presence of multiple decoupled morphological or
dynamical components, but rather from the fact that the instabilities
and resonances generated by the bar lead to a redistribution of the
disk material, and local modifications of its scalelength and
scaleheight.

Our simulations therefore suggest a close link between Freeman Type~II
profiles and bars, but this link is not entirely borne out by
observations. Although barred galaxies are more likely to harbor
Type~II profiles, bars appear insufficient. They are perhaps even
unnecessary. Indeed, not all face-on barred galaxies have type Type~II
profiles, and some non-barred galaxies do (e.g.\ \citealt*{bba96};
\citealt{mch03}). Type~II profile formation through bar-driven
evolution thus seems to require that at least some bars be destroyed
(while roughly maintaining their azimuthally-averaged profile). While
possible, the efficiency of such mechanisms is questionable (e.g.\
\citealt*{ss04,adl03} and references therein). The above observational
studies were based on optical studies, however, and it is known that
the fraction of barred disk galaxies increases in the near-infrared
(NIR; e.g.\ \citealt{sj98,eetal00}). It would thus be worthwhile to
redo the statistics with NIR data, and see if the correlation between
Type~II profiles and bars gets stronger.

In any case, while a proper comparison of our models with observations
is required, including full bulge-disk decompositions for a variety of
disk orientations, our model light profiles clearly support the
interpretation that many bulges are formed through bar-driven
rearrangement of disk material. The exact nature of Freeman Type~II
profiles remains to be determined.

\subsection{$h_3-V$ (Anti-)Correlation\label{sec:h3-v}}

It is likely that the kinematic bar signatures discussed for $I$, $V$,
and $\sigma$ can individually be created by carefully selected
axisymmetric density distributions or distribution functions. However,
whether this remains true when all the signatures are considered
simultaneously is unclear. Furthermore, the behavior of $h_3$ is
generally less well understood, particularly with respect to $V$, so
we consider some toy models below. Our goal is to understand under
which axisymmetric conditions an $h_3-V$ correlation can arise, and
thus invalidate its use as a triaxiality diagnostic.

We first consider a kinematically cold axisymmetric disk, and assume a
monotonically decreasing surface density profile. Such a disk can be
thought of as a sum of circular rings, each yielding a straight line
in an edge-on PVD (from the origin to the circular velocity at the
ring's radius). Because of the decreasing surface density, the LOSVD
at each (projected) position will systematically have a steep prograde
wing and a tail of low-velocity material, yielding the usual $h_3-V$
anti-correlation expected from an inclined disk. If the disk is
inwardly truncated, the rotation curve will be solid-body until the
inner ring's radius and then follow the circular velocity curve, and
the $h_3-V$ anti-correlation will be little affected. Adding an
outwardly truncated inner disk (i.e.\ creating a double-disk
structure), the velocity profile will become doubly peaked, but $h_3$
and $V$ will remain anti-correlated at all radii, with perhaps a break
at the end of the inner disk.

We can try relaxing the constraint on the surface density and consider
a disk which is not strictly inwardly truncated, but rather has a
lower surface density in the inner parts. Then, in the edge-on PVD,
the mean velocity will be dictated by the (projected) outer disk and
the inner disk will create a tail of high-velocity material, possibly
yielding a correlation of $h_3$ and $V$ without the need for
triaxiality (if the high velocity tail of the inner disk is more
important that the low-velocity one associated with the outer
disk). This situation is similar to that of a strong bar seen end-on,
as described in \S~\ref{sec:h3}, but there is no reason to expect such
an ab initio density profile in a kinematically cold axisymmetric disk
(although it can not be ruled out either). Truncated or double-disk
structures can however arise from the influence of a bar, and it is
often argued that their occurence in an axisymmetric galaxy indicates
the past presence of a bar long dissolved or destroyed \citep[see,
e.g.,][]{bba96,ebmp96,be98}. The same can thus perhaps be said of an
$h_3-V$ correlation.

A more interesting situation occurs if we consider an axisymmetric but
kinematically hot subsystems at the center of our cold disk. This
classical bulge-like component will have a mean velocity lower than
the circular velocity (because of pressure support), but if taken in
isolation will still yield an $h_3-V$ anti-correlation when viewed
edge-on. The same will be true if present within a truncated disk,
although the rotation curve will then likely be double-peaked. If this
classical bulge is inserted within a full disk, however, it is likely
that the exact behavior of $h_3$ will depend on the bulge-to-disk
ratio $B/D$. At any given (projected) position, the mean velocity
could be dictated by the classical bulge if it is bright enough, the
(projected) disk then creating a tail of high-velocity material, or
the inverse may be true if the bulge is faint. Both $h_3-V$
correlations and anti-correlations can thus probably occur as a
function of (projected) radius and $B/D$ (significant $h_4$ should
also be expected). In fact, \citet{mf93} showed that simple
spherically symmetric and isotropic models can yield both $h_3-V$
correlations and anti-correlations depending on the amount of
rotation. In particular, $h_3$ was found to correlate with $V$ for
maximally rotating models (see also \citealt{g93} for non-rotating
models).

It is thus clear from the above that an $h_3-V$ correlation is not
uniquely related to triaxiality. It is possible that it simply traces
a high surface brightness region with increased velocity dispersion
(i.e.\ a spheroid-like component) embedded within a disk, whether
triaxial or not. But it is unclear if any axisymmetric configuration
could simultaneously account for all of the kinematic bar signatures
observed (and their spatial correlations) in addition to $h_3$ (e.g.\
surface brightness plateau at moderate radii, double-hump rotation
curve, flat-top or slightly peaked velocity dispersion profile with a
secondary maximum and/or local central minimum, etc.). Although it is
beyond the scope of this paper to prove the contrary, this appears
rather unlikely since most of the features are directly related to the
particular orbital structure of barred disks. The kinematic bar
signatures discussed throughout this paper are thus not only
consistent with the presence of a bar, but they strongly argue for it.

Kinematic profiles derived from axisymmetric models with varying $B/D$
would be useful to clarify the relationship of $h_3$ with the
intrinsic shape of spheroids and the degree of pressure support. If
the $S/N$ of data is high enough, however, the PVDs themselves should
allow to break any degeneracy, since they will be significantly
different for the various configurations discussed (despite yielding
qualitatively similar Gauss-Hermite parameters). In future work, we
will study simulations which include a luminous axisymmetric bulge in
addition to the current luminous disk and dark halo. This should allow
us to quantify the effects of having a classical bulge in addition to
a barred disk.

\subsection{Central Velocity Dispersion Minima\label{sec:sig_min}}

As noted in \S~\ref{sec:sigma} and visible in
Figures~\ref{fig:strength}--\ref{fig:incl}, a central velocity
dispersion minimum is often present for strong bars seen approximately
end-on. This is very interesting since central $\sigma$ minima are
generally associated with fast rotating (i.e.\ kinematically cold)
disks, thought to form by gas inflow and subsequent star formation
(whether bar-driven or not; e.g.\ \citealt{hs94,fb95}). Here, however,
the central $\sigma$ minimum must arise from purely dissipationless
processes.

Although the axis ratio $a/b$ of the dominant $x_1$ orbits in bars
generally increases with decreasing radius, this behavior is inverted
near the center where the orbits become progressively more circular
(although they can still be very elongated; see, e.g., \citealt{a92a};
\citeauthor{ba99}). This property is probably the main driving
mechanism behind the central $\sigma$ minima observed in our
simulations. The relative shallowness of the minima is likely due to
the facts that the bar density is not homogeneous in the inner parts
and the axial ratio need not tend to $1$ until very close to the
center. The line-of-sight integration through the outer parts of the
bar and disk must of course also be considered.

Although the central $\sigma$ minimum for the strong bar case
discussed here is comparable to those observed in the sample of
\citet{cb04}, it is probably shallower than the deepest central
stellar velocity dispersion minima known (e.g.\ \objectname{NGC~6503},
\citealt{b89} and \citealt{bg97}; \objectname{NGC~1365},
\citealt{agcflpw01}). While gaseous processes clearly can not be
neglected in those cases, we recall that we have not attempted to fit
any particular object, and other stellar processes can also increase a
purely stellar central $\sigma$ minimum. For example, the central
density profile can be less peaked (i.e.\ more homogeneous) and the
bar (and orbits) axial ratio can decrease more rapidly in the inner
parts. This is particularly likely as the central $a/b$ profile of the
simulations is generally flatter than in real bars
\citep[see][]{am02,gs03}. Gas inflow will futher help, first by
circularizing the potential in the inner parts, and second by adding a
kinematically cold population of young stars if star formation occurs.

\subsection{Observations\label{sec:observations}}

\citet{cb04} provide an in-depth comparison of theirs and other
observations with our simulations, but we summarize here some
important facts. \citet{cb04} obtained optical stellar kinematics
along the major-axis for a sample of $30$ edge-on spiral galaxies
(S0--Sbc), $80\%$ of which have a B/PS bulge. For essentially all B/PS
bulges where the data are good enough, the predicted kinematic
signatures are observed: double-hump rotation curves with a dip or
plateau at moderate radii, often flat-top velocity dispersion profiles
with a secondary maximum or shoulder, and $h_3$ profiles correlated
with $V$ over the expected bar length. Examples include the
well-studied galaxies \objectname{NGC~128}, \objectname{NGC~1381}, and
\objectname{IC~4767}, all perfectly edge-on and largely dust-free S0s
with a B/PS bulge, as well as the later type spirals
\objectname{NGC~5746}, \objectname{NGC~6722}, and
\objectname{NGC~6771}. Furthermore, $40\%$ of the galaxies have a
local central $\sigma$ minimum, and essentially no bar signature is
observed in the non-B/PS bulges. The light profiles show a variety of
shapes, but this is not surprising given that many galaxies are
dusty. The $K$-band data of \citet{baabdf05} will yield a more useful
comparison \citep[see also][]{aabbdvp03}.

\citet{f97} also present the stellar kinematics of a number of edge-on
galaxies. Again, all B/PS bulges show the expected kinematic
signatures. Two non-B/PS bulges also do, but they are consistent with
bars seen end-on. The observations of \objectname{NGC~4762} are
particularly interesting in this respect. \objectname{NGC~4762} is an
edge-on S0 galaxy with a small round bulge and at least two extended
plateaus in its major-axis surface brightness profile
\citep{b79,t80,wh84}. It however appears very thin over the first
plateau, traditionally associated with a bar, which has prompted
speculation that either the bar is thin (i.e.\ it has not buckled) or
the plateau is not related to a bar at all
\citep[e.g.][]{w94}. However, within the radial range probed by our
simulations, both the morphology, photometry, and stellar kinematics
are perfectly consistent with that expected for an end-on bar. In
fact, \citeauthor{f97}'s (\citeyear{f97}) data are almost a perfect
match to our generic simulation of a strong bar seen end-on, as shown
in Figure~\ref{fig:strength}. Similar cases include
\objectname{NGC~4350} \citep{f97} and \objectname{IC~5096}
\citep{cb04}.

A number of the (presumably) barred galaxies in the \citet{f97} and
\citet{cb04} samples also show an $h_3-V$ anti-correlation in the very
center, contrary to what is observed here. This implies the presence
of an additional central rapidly rotating component. Given that many
of those galaxies also possess a gaseous nuclear spiral \citep{bf99},
it is natural to postulate that this component arises from a younger
and kinematically cold stellar population, presumably formed through
bar-driven inflow and star formation
\citep[e.g.][]{hs94,fb93,fb95}. Our simulations simply do not take
into account this potential dissipative component. As discussed above
(\S~\ref{sec:sig_min}), it will contribute further to the central
$\sigma$ minimum observed in some galaxies.

The stellar kinematic observations of \citet{f97} and \citet{cb04} are
thus entirely consistent with our simulations and support the
interpretation that B/PS bulges are indeed thick bars. Interestingly,
signatures similar to the ones predicted here are also observed in
more face-on barred systems, as illustrated by the double-hump
rotation curves and central $\sigma$ minima observed by
\citet{agcflpw01} in $4$ moderately inclined intermediate-type
spirals. 2D kinematic fields from wide-field integral-field
spectrographs such as SAURON \citep{betal01} will provide the next
level of tests, along with a full 2D analysis of N-body simulations
(including higher order LOSVD terms), but preliminary results support
both the above long-slit work and our current simulations
\citep[e.g.][]{zetal02,betal02a,betal02b,eetal04}. Improved
morphological studies should also follow, and those will be provided
by the $K$-band observations of the \citet{cb04} sample
\citep{baabdf05,aab05} and by parallel work on the current
simulations.
%
%
\section{Conclusions\label{sec:conclusions}}
To constrain bar-driven evolution scenarios for the formation of
bulges, and more specifically to provide tools to probe the nature of
boxy and peanut-shaped (B/PS) bulges, we have used self-consistent 3D
N-body simulations of bar-unstable disks to study the kinematic
signatures of edge-on bars. Quantifying the major-axis stellar
kinematics with Gauss-Hermite polynomials, a number of features can be
used as bar diagnostics: 1) a steep quasi-exponential central light
profile with a shoulder or plateau at moderate radii; 2) a double-hump
rotation curve, possibly showing a local maximum and minimum at
moderate radii; 3) a sometime broad central velocity dispersion peak
with a shoulder or plateau (and possibly a secondary maximum) at
moderate radii; 4) line-of-sight velocity distributions (LOSVDs) with
a high-velocity tail (i.e.\ an $h_3-V$ correlation) over the projected
bar length. A local central $\sigma$ minimum can also be present for
strong bars seen approximately end-on. The positions and lengths of
those features are further correlated, providing an easy and reliable
tool to identify bars in edge-on disks. We note that despite previous
claims, a ``figure-of-eight'' is never seen in stellar
position-velocity diagrams (PVDs), as expected from realistic models
and orbital configurations.

Interestingly, while some of the kinematic bar features can
individually be created by axisymmetric density distributions, the
$h_3-V$ correlation appears to be a particularly reliable tracer of
triaxiality (although it is not uniquely related to it). A number of
those kinematic bar signatures can also be used as proxies for the bar
length, and thus as an indirect measure of the bar pattern speed. The
sharpness of the kinematic features generally decreases for weaker bar
and/or increasing viewing angle (from end-on to side-on), introducing
some degeneracy between the two but ensuring complementary with
morphological bar signatures (strongest for side-on bars). All
characteristic features of the kinematic profiles are established the
moment the bar forms and grow stronger with time, as the bar lengthens
and strengthens. They vary little for small inclination variations but
noticeable differences appear for $i\lesssim80\degr$.

Existing data and the recent work of \citet{cb04} show that most B/PS
bulges indeed show the kinematic signatures identified here, lending
support to evolution scenarios where those bulges are formed through
vertical disk instabilities. Detailed work on the morphological and
photometric properties of both simulated and observed B/PS bulges are
ongoing \citep{baabdf05,aab05} and should provide more tests of the
consistency of these scenarios. Preliminary results support our
current conclusions.

The current simulations also systematically produce Freeman Type~II
(truncated) surface brightness profiles with approximately exponential
peaks, as is observed in most bulges (at least late-types). The local
central $\sigma$ minimum present in the strongest bars (and produced
without dissipation) is also increasingly thought to be common in
spirals.

\acknowledgements Support for this work was provided by NASA through
Hubble Fellowship grant HST-HF-01136.01 awarded by the Space Telescope
Science Institute, which is operated by the Association of
Universities for Research in Astronomy, Inc., for NASA, under contract
NAS~5-26555. EA thanks J.~C.\ Lambert for his help with the GRAPE
software and the administration of the simulations. EA also thanks the
INSU/CNRS, the University of Aix-Marseille~I, the Region PACA, and the
IGRAP for funds to develop the GRAPE and Beowulf computing facilities
used for the simulations and their analysis. We wish to thank K.\ C.\
Freeman for his involvement in the early stages of this project as
well A.\ Bosma, A.\ Chung, K.\ Kuijken, and R.\ van der Marel for
useful discussions.
%
%

%
\clearpage
%
%
%
\figcaption[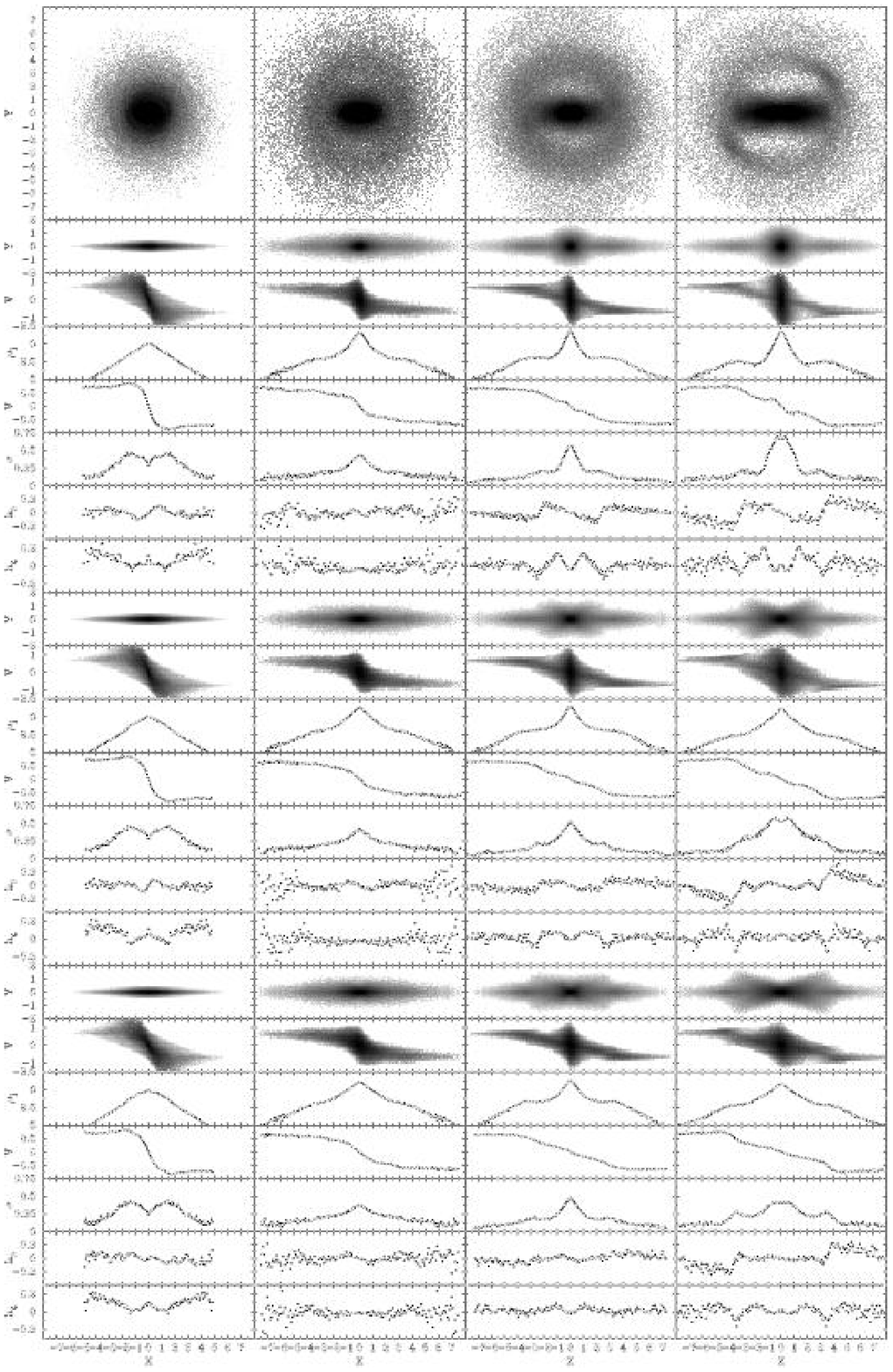]{Bar diagnostics as a function of viewing angle and
  bar strength. From left to right: no bar, weak bar, intermediate
  bar, and strong bar case. From top to bottom: face-view of the
  simulation and stellar kinematics for the bar seen end-on
  ($\psi=0\degr$), at intermediate angle ($\psi=45\degr$), and side-on
  ($\psi=90\degr$). From top to bottom, each kinematic panel shows an
  edge-on view of the simulation, the extracted PVD along the
  major-axis, the major-axis surface brightness profile ($\mu_I$), as
  well as the derived Gauss-Hermite coefficients $V$ (mean velocity),
  $\sigma$ (velocity dispersion), $h_3$ (skewness), and $h_4$
  (kurtosis). Because of strong gradients, all grayscales are plotted
  on a logarithmic scale.\label{fig:strength}}
%
%
\figcaption[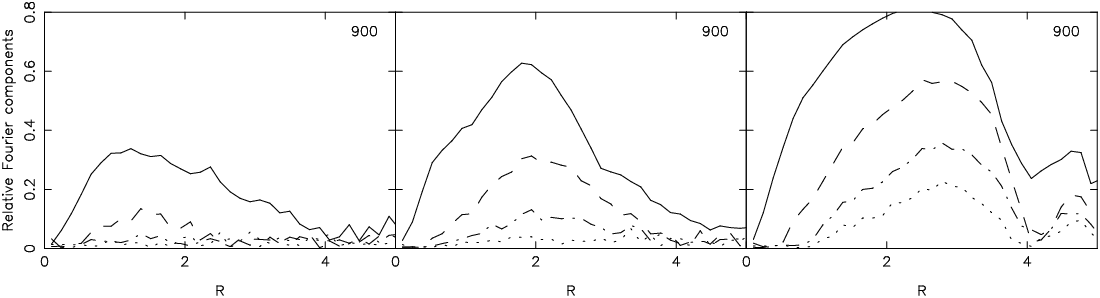]{Fourier decompositions of the face-on density
  distributions as a function of bar strength
  \citep[see][]{am02}. From left to right, the relative importance of
  the even Fourier terms (compared to the $m=0$ axisymmetric term)
  are shown for the weakly, intermediate, and strongly barred
  models. The terms shown are $m=2$ (solid line), $m=4$ (dashed line),
  $m=6$ (dot-dashed line), and $m=8$ (dotted
  line).\label{fig:fourier}}
%
%
\figcaption[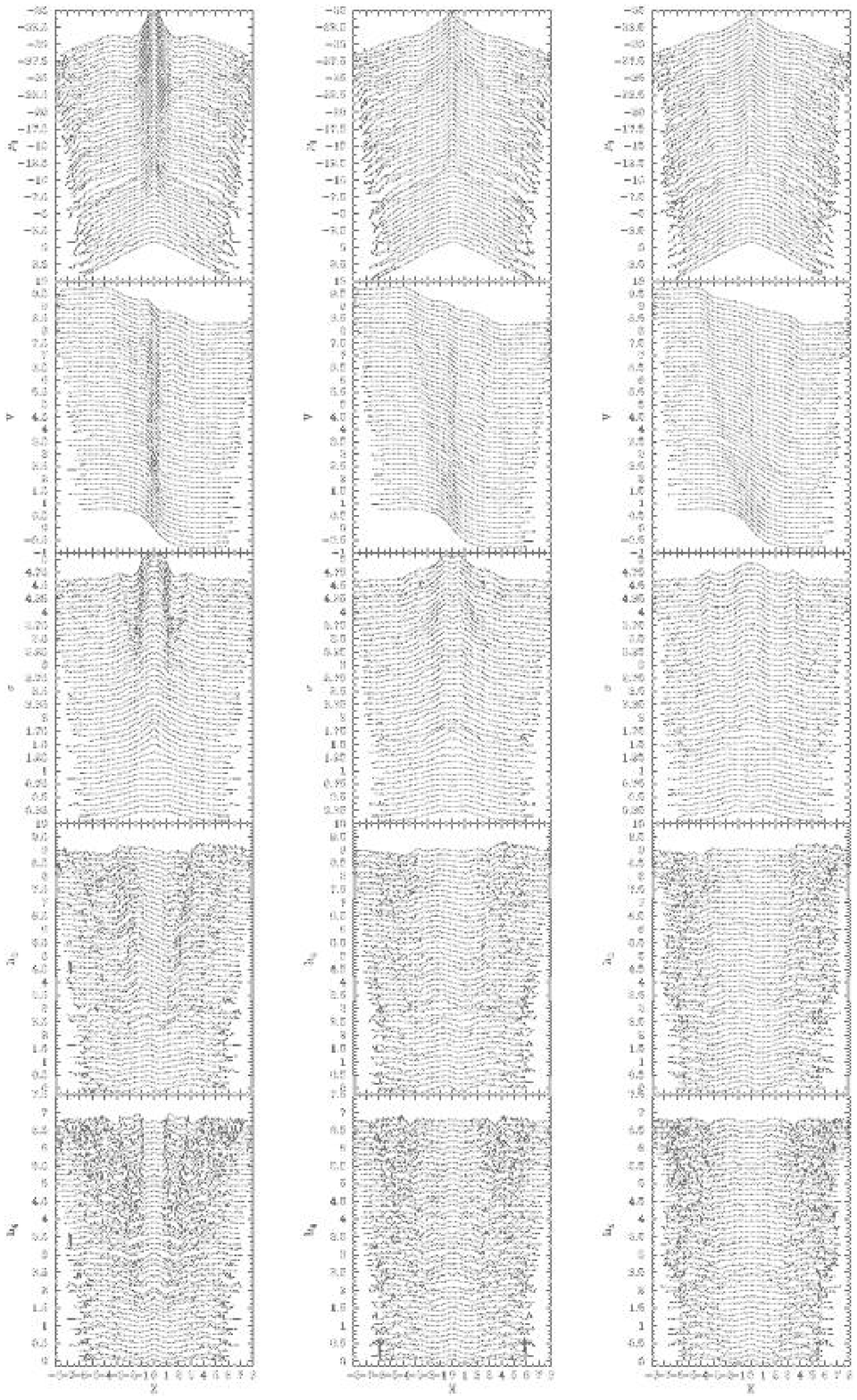]{Bar diagnostics as a function of viewing angle and
  time for the strong bar case. From top to bottom: the evolution of
  the major-axis surface brightness profile ($\mu_I$) and the derived
  Gauss-Hermite coefficients $V$ (mean velocity), $\sigma$ (velocity
  dispersion), $h_3$ (skewness), and $h_4$ (kurtosis). From left to
  right: stellar kinematics for the bar seen end-on ($\psi=0\degr$),
  at intermediate angle ($\psi=45\degr$), and side-on
  ($\psi=90\degr$). Each kinematic panel shows the time evolution of a
  profile in steps of $20$ time units. $t=0$ is at the bottom and the
  profiles are offset vertically from each other by an arbitrary
  amount.\label{fig:time}}
%
%
\figcaption[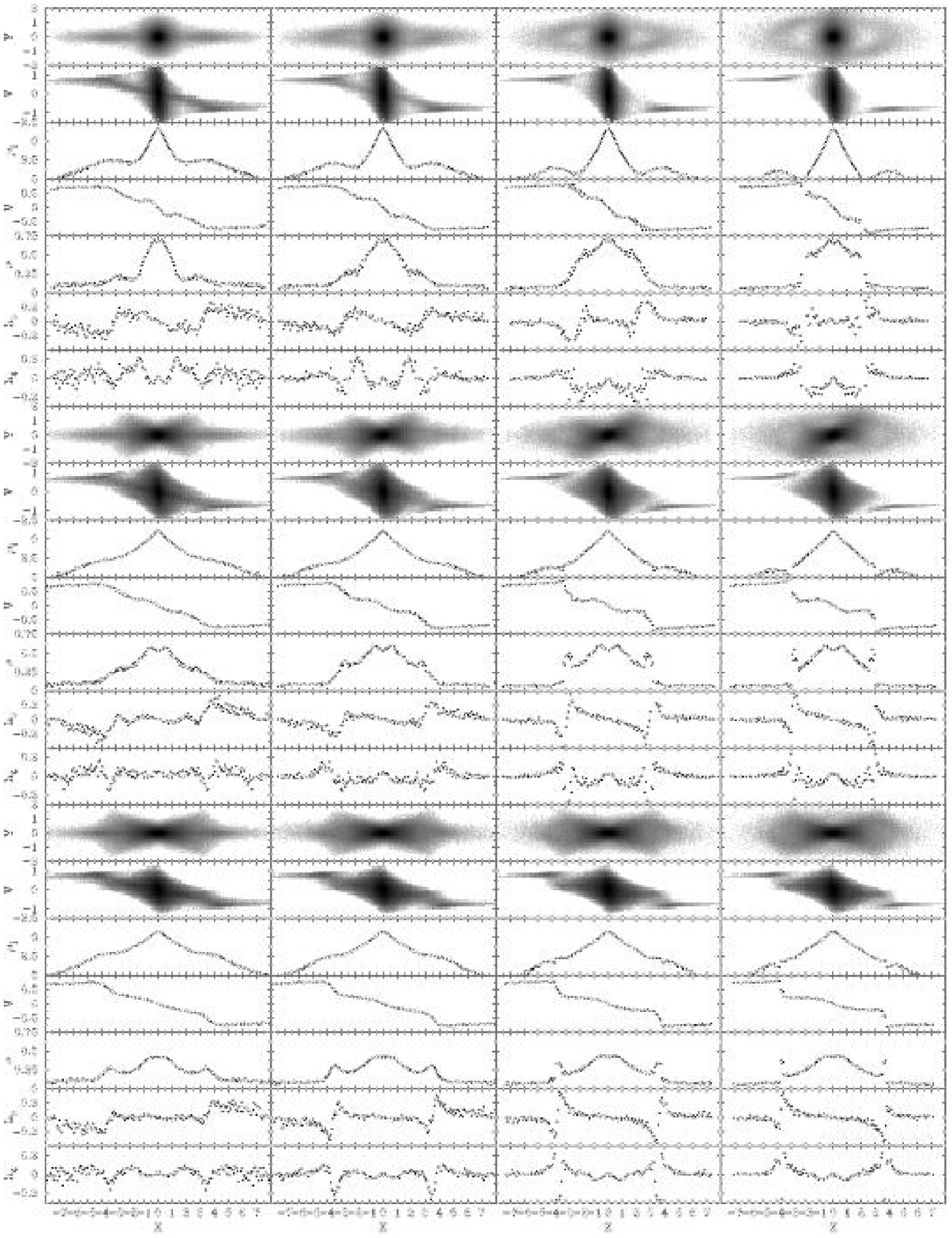]{Bar diagnostics as a function of viewing angle and
  inclination for the strong bar case. From left to right: $i=90$, $85$, $80$,
  and $75\degr$. From top to bottom: stellar kinematics for the bar seen
  end-on ($\psi=0\degr$), at intermediate angle ($\psi=45\degr$), and side-on
  ($\psi=90\degr$). From top to bottom, each kinematic panel shows a properly
  projected view of the simulation, the extracted PVD along the major-axis,
  the major-axis surface brightness profile ($\mu_I$), as well as the derived
  Gauss-Hermite coefficients $V$ (mean velocity), $\sigma$ (velocity
  dispersion), $h_3$ (skewness), and $h_4$ (kurtosis). Because of strong
  gradients, all grayscales are plotted on a logarithmic
  scale.\label{fig:incl}}
%
%
%
\begin{figure}
\epsscale{0.84}
\plotone{f1.eps}
\epsscale{1.0}
\end{figure}
\clearpage
\begin{figure}
\plotone{f2.eps}
\end{figure}
\clearpage
\begin{figure}
\epsscale{0.79}
\plotone{f3.eps}
\epsscale{1.0}
\end{figure}
\clearpage
\begin{figure}
\plotone{f4.eps}
\end{figure}
\end{document}